# Prosthetic finger phalanges with lifelike skin compliance for low-force social touching interactions

John-John Cabibihan*, Raditya Pradipta, Shuzhi Sam Ge

The Social Robotics Laboratory, Interactive and Digital Media Institute and

Department of Electrical and Computer Engineering,

The National University of Singapore, Singapore

*Corresponding author

Email addresses:

      JJC: elecjj@nus.edu.sg

      RP: raditya.pradipta@nus.edu.sg

      SSG: samge@nus.edu.sg

# Abstract

**Background**

Prosthetic arms and hands that can be controlled by the user's electromyography (EMG) signals are emerging. Eventually, these advanced prosthetic devices will be expected to touch and be touched by other people. As realistic as they may look, the currently available prosthetic hands have physical properties that are still far from the characteristics of human skins because they are much stiffer. In this paper, different configurations of synthetic finger phalanges have been investigated for their skin compliance behaviour and have been compared with the phalanges of the human fingers and a phalanx from a commercially available prosthetic hand.

**Methods**

Handshake tests were performed to identify which areas on the human hand experience high contact forces. After these areas were determined, experiments were done on selected areas using an indenting probe to obtain the force-displacement curves. Finite element simulations were used to compare the force-displacement results of the synthetic finger phalanx designs with that of the experimental results from the human and prosthetic finger phalanges. The simulation models were used to investigate the effects of (i) varying the internal topology of the finger phalanx and (ii) varying different materials for the internal and external layers.

**Results and Conclusions**

During handshake, the high magnitudes of contact forces were observed at the areas where the full grasping enclosure of the other person's hand can be achieved. From these areas, the middle phalanges of the (a) little, (b) ring, and (c) middle fingers were selected. The indentation experiments on these areas showed that a 2 N force corresponds to skin tissue displacements of more than 2 mm. The results from the simulation model show that introducing an open pocket with 2 mm height on the internal structure of synthetic finger phalanges increased the skin compliance of the silicone material to 235% and the polyurethane material to 436%, as compared to a configuration with a solid internal geometry. In addition, the study shows that an indentation of 2 N force on the synthetic skin with an open pocket can also achieve a displacement of more than 2 mm, while the finger phalanx from a commercially available prosthetic hand can only achieve 0.2 mm.

# Background

It is not very often that our days pass by without any of the following touch-related experience: a handshake from a colleague, a caress from a spouse, a hug, a pat on the back or high fives with pals. As social beings, the need to touch and be touched is inherent among us. Indeed, it is through touch that distinct emotions such as anger, fear, disgust, love, gratitude and sympathy can be communicated [1]. However, these typical exchanges of social touching may be limited for those who have lost their hands due to an accident, disease or war.

Prosthetic arms and hands that can be controlled by the electromyography (EMG) signals are emerging [2, 3]. Kuiken et al [4] demonstrated a surgical technique called targeted muscle reinnervation that transfers residual arm nerves to alternative muscle sites. They have demonstrated that this technique enables the subjects to perform real-time control of motorized shoulders, elbows, wrists and hands for grasping soft and brittle objects, like grapes or eggs, with enough force to hold them without crushing them. On the commercial front, Touch Bionics' lightweight and fully articulated prosthetic hand, the i-LIMB Hand [5], can now enable the amputees to perform gestures and various gripping movements (e.g. power, precision, index point, etc).

Eventually, these advanced prosthetic devices will be expected to perform affective touches on other people. This paper focuses on replicating the softness of the skin tissue on the human hand during social touching interactions. Particularly, different configurations of synthetic finger phalanges were investigated with the goal to replicate the softness of the human finger phalanges that is felt during handshakes and light touches. Here, softness is defined as the perceptual correlate of skin compliance or the amount of deformation caused by an applied force [6].

Aside from the functional limitations, the loss of hands implies negative psychological and emotional consequences on the person and to those around him or her. Using focus group methodology on 14 participants who have lost a limb, Gallagher and MacLachlan [7] found that self-image, social, physical and practical concerns are some of the key factors that are important in the adjustment process. They revealed that a common response in seeing the prosthesis for the first time was "extreme shock and disappointment". On the question of "What do you find most difficult to deal with having an artificial limb?" some participants expressed concern about the impression they made on others. One woman described that even in her own home she found it impossible for her to relax without her artificial limb on, anticipating that an unexpected visitor may arrive. Gallagher and MacLachlan noted that the common sentiment of the participants is to appear and be 'normal' again. With regard to social interactions, many of the participants encountered uncomfortable situations, wherein they stated that other people's reactions varied from patronization to complete shock.

Under similar circumstances and with limited alternatives for replacement limbs, it is understandable that depression can be one of the conditions associated with limb loss. Gallagher and MacLachlan [8] found that the rates of clinical depression range from 21 to 35 percent. Rybarczyk et al [9] highlighted five key issues that clinicians should attend to: 1) amputation is a diverse disability; 2) discrimination by others; 3) self-stigma; 4) feeling vulnerable to victimization; and 5) the role of values, meaning and perspective in positive adjustment. Using semi-structured email and face-to-face interviews on 35 prosthesis users, Murray [10] concluded that prosthesis use plays a social role in the lives of people with limb loss or absence. He found that the ability to conceal such use enabled participants to ward off social stigmatisation that could result into better integration into the society and the reduction of emotional problems.

Given the many issues surrounding limb loss or absence, many researchers have been addressing the technical issues in sensing, control and functionality of prosthetic arms and hands with some success (e.g. [11-16]). In terms of appearance, the current upper limb prosthetics can now be created to be indistinguishable from the natural ones with the accurate mimicry of the skin tone, hairs and pores (e.g. [17, 18]). However, these are still stiff when touched. The availability of artificial hands with lifelike appearance but not lifelike softness poses a perception problem to the person that these hands will touch. As such, finding synthetic materials that have similar properties to that of the human skin becomes important [19-21].

Among the social touches mentioned, the handshake remains one of the more acceptable gestures in many cultures. We conducted experiments to determine the areas of frequent contact during handshakes and the corresponding forces. After determining these areas, another set of experiments were done to indent a probe on the hands to obtain the amount of displacement that results from the application of forces during handshake contact. We then used Finite Element (FE) simulations to find synthetic skin designs that have similar force-displacement (i.e. skin compliance) characteristics to the human hand. The model was used to investigate the effects of (i) varying the internal topology of the finger phalanx and (ii) varying different materials for the internal and external layers. Comparisons were done on the skin compliance behaviours of the finger phalanges of the human hand, the phalanx of a commercially available prosthetic hand and the synthetic skin design with open pockets.

# Materials and Methods

## Handshake experiments

Handshake experiments were performed to investigate the typical range of contact forces and to find out which areas on the human hand have high contact forces during handshake. The experiments were limited to male-to-male handshake in light of the earlier findings in [22] that handshakes are more frequent in male-to-male pairs than in female-to-female or male-to-female pairs. Tactile force sensors were attached to the right hand of the male experimenter before he performed a series of handshakes with the test subjects. The force sensors (FingerTPS, Pressure Profile Systems, USA) are made of tactile pads that detect changes in capacitance when the upper and lower tactile pads come into contact.

Upon shaking hand with the test subjects, the forces exerted on the experimenter's hand make these pads touch each other, resulting into changes in the sensors' capacitance values. It would then be possible to measure the magnitude of the exerted forces. The sensors were placed on eighteen areas on the experimenter's hand. These included all the phalanges of the fingers, the two areas on the palm and the two areas on the back of the palm. As confirmed by pre-tests, these were the areas where contact can occur during handshake. The locations of the sensors and their naming conventions are shown in Fig. 1.

There were 30 male test subjects who participated in this handshake experiment. All of them were students or researchers at the National University of Singapore (NUS). Each of the test subjects were asked to perform a casual handshake with the experimenter. Prior to experiment, the experimenter was trained to give a neutral handshake, in which he kept his grasping force at a minimum and waited for the handshake partner to initiate the contact. A similar protocol of having a neutral handshake early in the sequence was carried out by Chaplin et al [23] to establish the relationship between handshakes and personality. The data on the experimenter and the experimental subjects are shown in Table 1. The test subjects were reimbursed for their participation. Approval for the handshake experimental protocol was granted by the NUS Institutional Review Board.

**Indentation experiments on the human and prosthetic hands**

In order to obtain the data to compare our simulation results with, indentation experiments were done on the phalanges where high contact forces occurred as revealed by the handshake experiments. The middle phalanges of the little, ring and middle fingers were chosen as the target sites. The choice was made for two reasons. First, these phalanges have similar functional roles in the context of the handshake; they support the lower part of the other person's hand during the grasp. Second, these phalanges have geometrical similarities and modelling them can be done with relative ease. We were also interested in seeing whether there are significant differences on the force-displacement data when the hand lies in a flat position (Fig. 2a) and when the fingers are in a curled position (Fig. 2b). A hand in a flat position can represent light touches to another person while fingers in a curled position can represent the hand orientation during handshake. Additionally, we wanted to know how the force-displacement data from the phalanx of a commercially available prosthetic hand compare with the human skin and the synthetic skin that we are investigating. We obtained a product sample of a prosthetic hand (Silicone Cosmetic Hand Model 102L, Regal Prosthesis Ltd, Hong Kong) being sold at a local prosthetics shop. We indented this prosthetic hand at the same locations of interest (Fig. 2c).

There were 10 male test subjects who participated in the hand indentation experiment. The data of these subjects are shown in Table 2. A testing machine (MicroTester™, Instron, UK), with a load cell limit of 5 N, was employed to make indentations on the finger phalanges. A specially fabricated brass indenter, with an indenting area of 20 mm x 10 mm, was slotted into the load cell. For the flat-hand position, the hand was positioned above a mould with the palm facing upwards. For the curled-fingers position, the hand was placed on a mould that can constrain the hand in a typical handshake position (Fig. 2b Inset). The mould was prepared with crystalline-silica free alginate (Alja-Safe®, Smooth-On Inc., USA). The surface of the finger phalanx of interest was placed in a normal position under the indenter. The indenter was then lowered until the force read-out from the testing machine reached approximately 0.05 N. This is negligibly small but is sufficient enough to verify that the indenter had contacted the finger phalanx. For each subject, all the test areas were indented with a ramp speed of 0.5 mm/s under a force of up to 2 N for the middle phalanges of the little, ring and middle fingers. Approval for the indentation experimental protocol was similarly granted by the NUS Institutional Review Board.

**Finite element modelling**

This section is composed of three sub-sections that describe the synthetic skin samples, the constitutive equations and the numerical simulations.

**Synthetic skin samples**

The skin materials for prosthetic and robotic fingers in [11, 24] were selected for this paper. Samples of silicone (GLS 40, Prochima, s.n.c., Italy) and polyurethane (Poly 74-45, Polytek Devt Corp, USA) were previously characterized in [19] for their viscoelastic and hyperelastic behaviours. The silicone sample has a Shore A value of 11 while the polyurethane sample has a value of 45. A lower value indicates a low resistance to an indenter in a standard durometer test. From the durometer values above, the selected silicone material is softer as compared to the polyurethane material.

**Constitutive equations**

The synthetic skins were assumed to behave with hyperelastic and viscoelastic properties. As such, the total stress was made equivalent to the sum of the hyperelastic (HE) stress and the viscoelastic (VE) stress such that:

$$\sigma(t) = \sigma_{HE}(t) + \sigma_{VE}(t) \tag{1}$$

where $t$ is the time. A strain energy function, $U$, defined in Storakers [25] for highly compressible elastomers was used to describe the hyperelastic behaviour of the synthetic materials. This has been found to achieve good fits on the experimental data of synthetic materials [19] and it was likewise implemented in a human fingertip model in [26]. The function is given as:

$$U = \sum_{i=1}^{N} \frac{2\mu_i}{\alpha_i^2}\left[ \lambda_1^{\alpha_i} + \lambda_2^{\alpha_i} + \lambda_3^{\alpha_i} - 3 + \frac{1}{\beta}(J^{-\alpha_i\beta} - 1) \right] \qquad (2)$$

where $\mu_i$ denote the shear moduli, $\alpha_i$ are dimensionless material parameters, $\lambda_i$ are the principal stretch ratios, $J = \lambda_1\lambda_2\lambda_3$ is the volume ratio and $N$ is the number of terms used in the strain energy function. The coefficient $\beta$ determines the degrees of compressibility in the energy function. The relationship of $\beta$ to the Poisson's ratio, $\upsilon$, is $\beta = \upsilon/(1-2\upsilon)$.

The hyperelastic stress is related to the strain energy function (2) by:

$$\sigma_{HE} = \frac{2}{J} F \frac{\partial U}{\partial C} F^{\mathrm{T}} \qquad (3)$$

where $F$ is the deformation gradient and $C$ is the right Cauchy-Green deformation tensors.

The viscoelastic behaviour is defined below, with a relaxation function $g(t)$ applied to the hyperelastic stress:

$$\sigma_{VE} = \int_0^t \dot{g}(\tau)\sigma_{HE}(t-\tau)\,d\tau \qquad (4)$$

The viscoelastic material is defined by a Prony series expansion of the relaxation function [27]:

$$g(t) = \left[ 1 - \sum_{i=1}^{N_G} g_i (1 - e^{-t/\tau_i}) \right]$$

(5)

where $g_i$ is the shear relaxation modulus ratio, $\tau_i$ is the relaxation time, and $N_G$ denotes the number of terms used in the relaxation function. The detailed information on how the governing equations are numerically solved have been described in the Abaqus/CAE Theory Manual [28].

Table 3 shows the material parameters for silicone and polyurethane. These material parameters were validated in [19]. The validation procedures in that paper consisted of having the indentation results in the finite element models matched against the results of the physical samples of synthetic fingers that were made of silicone and polyurethane materials. The results from simulation and validation experiments were in good agreement.

**The FE Model and Numerical Simulations**

Simulations were conducted to determine the effects of varying (i) the internal topology and (ii) varying the material combinations of the layers in the skin compliance result of the synthetic finger phalanges. The three-dimensional geometries of the finger phalanx designs are shown in Fig. 3. These were modelled using the commercial finite element analysis software Abaqus[TM] / Standard 6.8-EF (Dassault Systemes Simulia Corp., Providence, RI, USA). The simulations were run at the Supercomputing and Visualisation Unit of the Computer Centre at the National University of Singapore. The finger phalanx width is 16 mm, the height is 9 mm and the thickness is 10 mm. The internal layer was made to have three topologies: a solid internal geometry (Fig. 3a) and arc-shaped pockets with 1 mm (Fig. 3b) and 2 mm heights (Fig. 3c). Fig. 3d shows the detailed geometry consisting of two layers. The external layer has a 0.8 mm thickness, which was approximated to be the combined thickness of the epidermis and dermis skin layers of the human finger.

The effects of the different internal topologies are to be investigated with the use of the geometries given in Fig. 3. To investigate the effects of the different material combinations, the material coefficients (i.e. data in Table 3) of the external and internal layers were set in the Abaqus[TM] software. For example, to have a homogeneous solid material of silicone, the inner and outer layers were given the same set of material coefficients; to have silicone as the inner layer and polyurethane as the outer layer, the material coefficients were set accordingly.

Three sets of contact interactions were specified in the model. First, a 'normal' contact behaviour was applied on the surface of the indenting plate and the external surface of the finger phalanx model. Next, a tie-connection was assumed for the 0.8 mm external layer and the rest of the finger phalanx model. Lastly, a 'normal' contact behaviour was similarly applied on the upper and lower surfaces of the 1 mm and 2 mm pocket designs as they come into contact due to the indenting plate.

The Abaqus$^{TM}$ 6.8-EF tetrahedral elements were used in conjunction with its automatic seed mesh feature. There were 1260 elements automatically generated for the solid internal geometry, 4838 elements for the geometry with 1 mm pocket and 3290 elements for the geometry with 2 mm pocket. The base of the finger geometry was constrained in all degrees of freedom to represent the bone structure of the human finger.

A displacement loading condition was applied on the rigid analytical surface that progressively indented each of the finger phalanx designs. The loading rate was 0.5 mm/sec. The results corresponding to the normal force (i.e. RF2) and the vertical displacement (i.e. U2) were obtained. These results will be compared to the skin compliance data of the human finger phalanges and the prosthetic hand that were obtained from the indentation experiments.

## Results and Discussion

### Handshake experiments

The results of the handshake experiments are plotted in Fig. 4. The locations of the sensors on the hand are shown on the x-axis while the y-axis shows the force results from the tactile sensors. High contact forces (i.e. forces greater than 2 N) were experienced at the palm, back of the palm, the thumb, the proximal phalanx of the little finger and the middle phalanges of the little, ring and middle fingers. These phalanges are the locations where the full grasping enclosure of the other person's hand can be achieved. For the purpose of the indentation experiments, the middle phalanges of the little, ring and middle fingers were selected. These are henceforth named Little2, Ring2 and Middle2, respectively.

**Indentation experiments on the human hand**

The force–displacement curves obtained from the three test areas as well as the comparison of the skin tissue displacements at 2 N force for both the flat-hand position and the curled-fingers position are plotted on Fig. 5. The representative data from one subject in Fig. 5a, 5b and 5c show that indentation forces of 2 N can result into finger tissue displacements that can reach beyond 3 mm.

Fig. 5d shows the displacements of the finger phalanges of the 10 subjects at 2 N load under the curled and flat orientations. Paired-samples t-tests were conducted on the displacement data of each of these flat and curled pairs to evaluate the differences in their skin compliance results. There was no statistically significant difference found in the displacement data of the flat Little2 ($M = 3.2631$, $SD = 0.8426$) and curled Little2 ($M = 3.7862$, $SD = 1.1647$), t(9) = 1.6224, $p = 0.1392$ (two-tailed). The mean difference in the displacement data is 0.5231 with a 95% confidence interval ranging from -0.2063 to 1.2527.

No statistically significant difference was observed in the displacement data of the flat Ring2 ($M = 3.7163$, $SD = 0.4980$) and curled Ring2 ($M = 3.9892$, $SD = 1.3457$), t(9) = 0.7246, $p = 0.4871$ (two-tailed). The mean difference in the displacement data is 0.2729 with a 95% confidence interval ranging from -0.5789 to 1.1245.

Lastly, there was also no statistically significant difference in the displacement data in flat Middle2 ($M$ = 3.7452, $SD$ = 0.6623) and curled Middle2 ($M$ =3.5830, $SD$ = 0.7893), t(9) = -0.6169, $p$ = 0.5526 (two-tailed). The mean difference in the displacement data is 0.1622 with a 95% confidence interval ranging from -0.7573 to 0.4328. In summary, these data show that there are no significant differences in the skin compliance of the Little2, Ring2 and Middle2 in flat and curled orientations.

**Effect of the open pockets**

Fig. 6 shows the vertical displacement contours that correspond to the 2 N indentations for the solid internal geometry, the 1 mm and 2 mm height open pockets designs. The effect of having pockets on the finger phalanx models with a single material layer are shown by the thicker lines in Fig. 7 (i.e. Silicone (SIL) and Polyurethane (PU)). The simulation results of the models with solid internal geometry are shown at the bottom cluster in this figure. A 2 N compressive load resulted into 0.42 mm displacement for PU and a 0.77 mm displacement for SIL. Introducing a 1 mm height pocket increased the displacement to 1.38 mm for PU and 1.72 mm for SIL. These correspond to 229% and 123% increase in the displacement values, respectively, when compared to the solid internal geometry configuration. Having a 2 mm height pocket results into displacements of 2.25 mm for PU and 2.58 mm for SIL, corresponding to 436% and 235% increase, respectively, from the solid internal geometry configuration. These results show that having internal pockets can significantly increase the skin compliance results of synthetic skins.

**Effect of varying the layers**

The human skin is tough, compliant and has self-healing properties. Technologies that can replicate all these properties are not yet available for prosthetic skins. Therefore, it is important to investigate the effects of a two-layered synthetic skin, which can give insights on how to satisfy the requirements for softness for social touching interactions and other requirements for wear, puncture and tear.

This section describes the effects of having a 0.8 mm outer layer of one type of material and an internal layer of another material. The results are shown as the thinner line types in Fig. 7 (i.e. SIL Inner PU Outer and PU Inner SIL Outer). The first four curves clustered at the bottom part of the figure are the simulation results from the solid internal geometry configuration. A 2 N compressive load for a combination of PU inner layer and an outer layer of SIL resulted into a displacement of 0.44 mm, or a 4.8% increase from a homogeneous PU material condition. With a combined inner layer of SIL and outer layer of PU, the displacement was 0.72 mm or a 6.5% decrease from a 'SIL-only' material condition. These results were expected because the SIL material has a lower durometer value (i.e. softer) as compared to the PU material. For the remaining combinations, the changes in the displacements at the 2 N compressive load correspond to an increase or decrease of displacement values to within 7% from the homogeneous material condition.

The significant effect of having the two layers of materials can be observed during the unsupported deflection of the upper part of the pocket, where the effects of the material softness come in. Looking at the results of the internal pockets with 2 mm heights, we can observe that at the 2 mm displacement the magnitude of the force for the 'SIL-only' condition is 0.39 N and it is 0.5 N for the 'SIL inner and PU outer' configuration; or a 28% increase in force value. For the 'PU-only' configuration, the force is reduced from 0.84 N to 0.72 N for the 'PU inner and SIL outer' configuration, or about 14% decrease.

Alternatively, we can analyze the effects of having the two material layers by comparing the slopes (i.e. Δdisplacement/Δforce) of the rising part of the curves for the 1 mm and 2 mm internal pockets in Fig. 7. Again, for the geometries with the 2 mm height pockets as an example, the rising slope is 6.64 for the 'SIL-only' and 5.03 for 'SIL inner and PU outer' conditions. The slope is 3.25 for 'PU-only' and 4.29 for 'PU inner and SIL outer' conditions. Taken together, the 0.8 mm outer layer significantly affects the slope of the rising curve, which results into an increase or decrease of the force magnitude. These occur before the top layer of the pocket comes into contact with the bottom layer of the pocket and eventually stiffens.

**Comparisons of simulation results against the prosthetic and human finger phalanges**

The figures in the left column in Fig. 8 are plots of the resulting force-displacement curves from three sets of data. First, it shows the experimental results on the curled and flat human finger postures during indentation. These data are from Little2 (Fig. 8a), Ring2 (Fig. 8b), and Middle2 (Fig. 8c), which were chosen as the representative parts of the human hand that the experiments have shown to have high contact forces during handshake interactions. Second, the figure shows the indentation results on a finger phalanx of a prosthetic hand. Third, the simulation results of the finger phalanx design with 2 mm inner pockets are overlaid on the experimental results for comparison.

The figures in the right column in Fig. 8 show the magnitude of the displacements corresponding to the 1 N force indentation. The results labelled with 'curled' and 'flat' in Fig. 8 are from the experimental data taken from the finger phalanges indentation of the 10 test subjects. From the simulation results of the synthetic finger phalanges, it can be observed from the bars that having a 2 mm height internal pocket can introduce significant improvements in the skin compliance. Such results are important particularly when they are compared against the skin compliance of the finger phalanges of a commercially available prosthetic hand. The silicone material used by the manufacturer of the prosthetic hand was stiff as shown by the 0.2 mm deformation. Many prosthetic hands are being sold for their cosmetic appearance and durability. The development of prosthetic hands for natural social touching interactions has not been the norm.

## Conclusions

This paper addressed the continuing need for improved methods and designs that can make prosthetic hands and arms unnoticed during social touching situations. In addition to lifelike appearance, warmth and motion, prosthetic skins that can replicate the natural softness of the human hand may be able to shield the user from social stigma. This could lead to the faster improvement of his or her emotional well-being and permit the resumption of a normal life (cf. [10, 17, 29]).

In this paper, the areas of the hand where typical contact occurs during male-to-male handshakes were determined. The results show that the following areas of the hand have contact forces greater than 2 N when grasped during handshake: (a) the palm and the back of the palm, (b) the thumb, (c) the proximal phalanx of the little finger and the (e) middle phalanges of the little, (f) ring, and (g) middle fingers. These are the areas that envelope the handshaking partner's hand for a full grasp.

The middle phalanges of the little, ring and middle fingers were selected for indentations with a testing machine. The force-displacement curves were obtained on both the flat hand position, which represents tapping or caressing postures, and the curled-fingers position, which represents handshake postures. The indentation results show that the skin tissues at the finger phalanges are compliant and are exhibiting large displacements with minimal forces applied, i.e., a 1 N force corresponds to skin tissue displacements of more than 2 mm. The results also show that there are no significant differences in the force-displacement data on flat-hand position and curled-fingers position on the middle phalanges of the little, ring and middle fingers.

Three-dimensional finite element models were presented for investigating the effects of varying the internal topology and varying the material layers in an attempt to duplicate the skin compliance of human finger phalanges. The following conclusions can be made from the simulation results. First, the skin compliance can be increased by introducing open pockets on the internal structure of a synthetic finger phalanx. An arc-shaped pocket with a 2 mm height on the internal structure increased the skin compliance of the silicone material to as high as 235% and the polyurethane material to 436%, as compared to a configuration with a solid internal geometry.

Second, having one type of material for the 0.8 mm external layer and another type for the internal layer can affect the deflection of the finger phalanges' surface, but this combination has minimal effect when the top layer of the pocket comes into contact with the base of the finger phalanx. By knowing the effects of having multi-material layers, we can take advantage of a synthetic skin design with a stiff external layer and a soft internal layer. A stiff external layer can better protect the tactile sensors and electronics that may be embedded on the internal structure, while the soft internal layer can satisfy the requirements for more natural social touching.

Lastly, the simulation results show that the synthetic skins with the configurations described herein could achieve lifelike skin compliance for light social touches, especially under applied forces of about 1 N. The internal pockets can significantly improve the compliance of the synthetic skins that will be used for prosthetics. Future studies can investigate other softer materials (i.e. materials with lower Shore durometer values), find the optimal thickness for the internal and external layers and the internal pockets, and optimize the right combinations of materials to be used as the internal and the external layer.

# Competing interests

The authors declare that they have no competing interests.

# Authors' contributions

JJC designed the experiments, developed the simulations, performed the data analysis and contributed to the drafting of the manuscript. RP collected, processed and helped analyze the data. SSG participated in the design of the study, analysis of the data and contributed to the drafting of the manuscript. All authors have read and approved the manuscript. A preliminary version of this paper was earlier presented at the International Conference on Social Robotics at Incheon, Korea in 2009.

# Acknowledgements

This work was supported by the project 'Design of Prosthetic Skins with Humanlike Softness' (R-263-000-506-133) funded by the Academic Research Fund, Ministry of Education, Singapore. We thank Lifeforce Limbs and Rehab Pte Ltd for the prosthetic hand sample.

**Table 1** – Data for the Experimenter and Subjects (n = 30)

|  | Experimenter | Subjects (Mean ±Std Dev) |
|---|---|---|
| Age | 22 | 26.35 ± 3.36 |
| Height (cm) | 175 | 173.42 ± 4.37 |
| Weight (kg) | 70 | 65.65 ± 8.57 |
| Hand Length (mm) | 190 | 187.50 ± 11.37 |
| Hand Width (mm) | 88 | 89.53 ± 6.80 |

**Table 2** – Subjects' Data for the Indentation Experiment (n = 10)

|  | Subjects (Mean ±Std Dev) |
|---|---|
| Age | 26.11 ± 3.28 |
| Height (cm) | 173.40 ± 4.93 |
| Weight (kg) | 68.20 ± 18.23 |
| Hand Length (mm) | 188.10 ± 6.87 |
| Hand Width (mm) | 85.01 ± 8.49 |

**Table 3** – Coefficients for the Synthetic Materials

| $i$ | 1 | 2 | 3 |
|---|---|---|---|
| **Silicone ($\upsilon = 0.49$)** | | | |
| $g_i$ | 0.015 | 0.044 | 0.029 |
| $\tau_i$ (sec) | 0.025 | 0.150 | 0.300 |
| $\mu_i$ (MPa) | 0.080 | 0.010 | - |
| $\alpha_i$ | 0.001 | 15.500 | - |
| **Polyurethane ($\upsilon = 0.47$)** | | | |
| $g_i$ | 0.167 | 0.158 | 0.113 |
| $\tau_i$ (sec) | 0.100 | 1.380 | 25.472 |
| $\mu_i$ (MPa) | 0.100 | 0.063 | - |
| $\alpha_i$ | 5.500 | 8.250 | - |

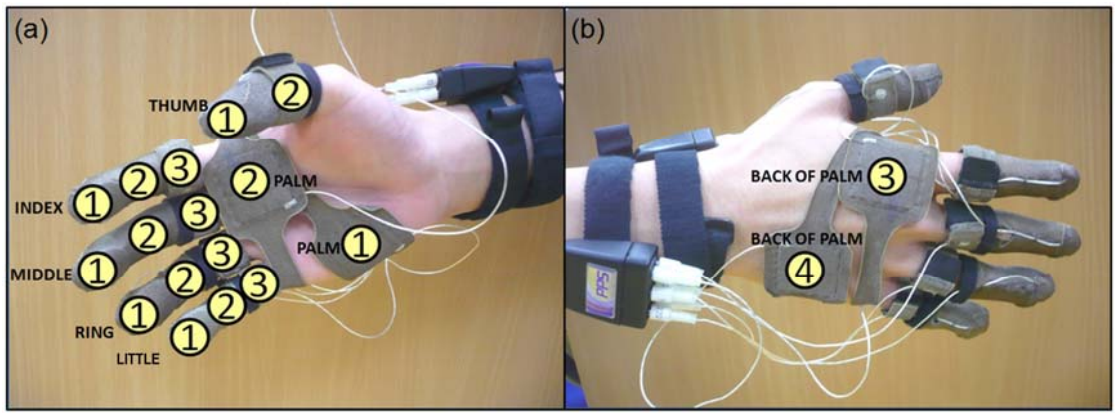

**Figure 1** – Location and naming convention of the force sensors on the hand (a) On the surface of the palm and (b) Back of the hand.

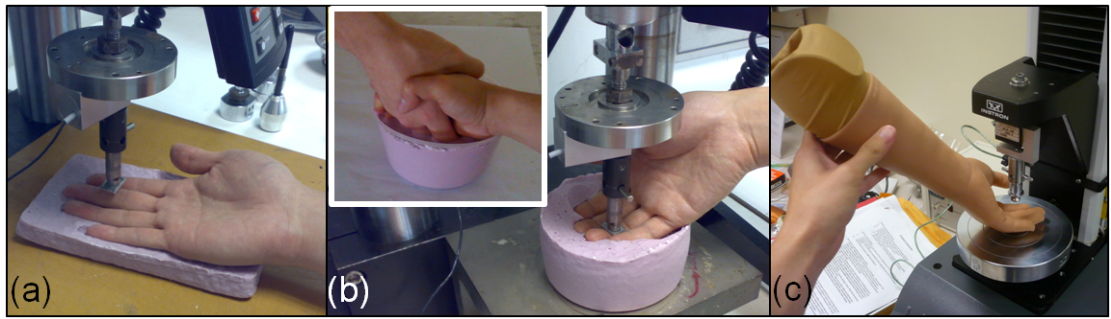

**Figure 2** – Setup for the indentation experiments (a) Flat-hand position, (b) Curled-fingers position, and (c) Prosthetic hand indentation. The Inset in (b) shows how the mould was prepared to restrict the hand in a typical handshake posture.

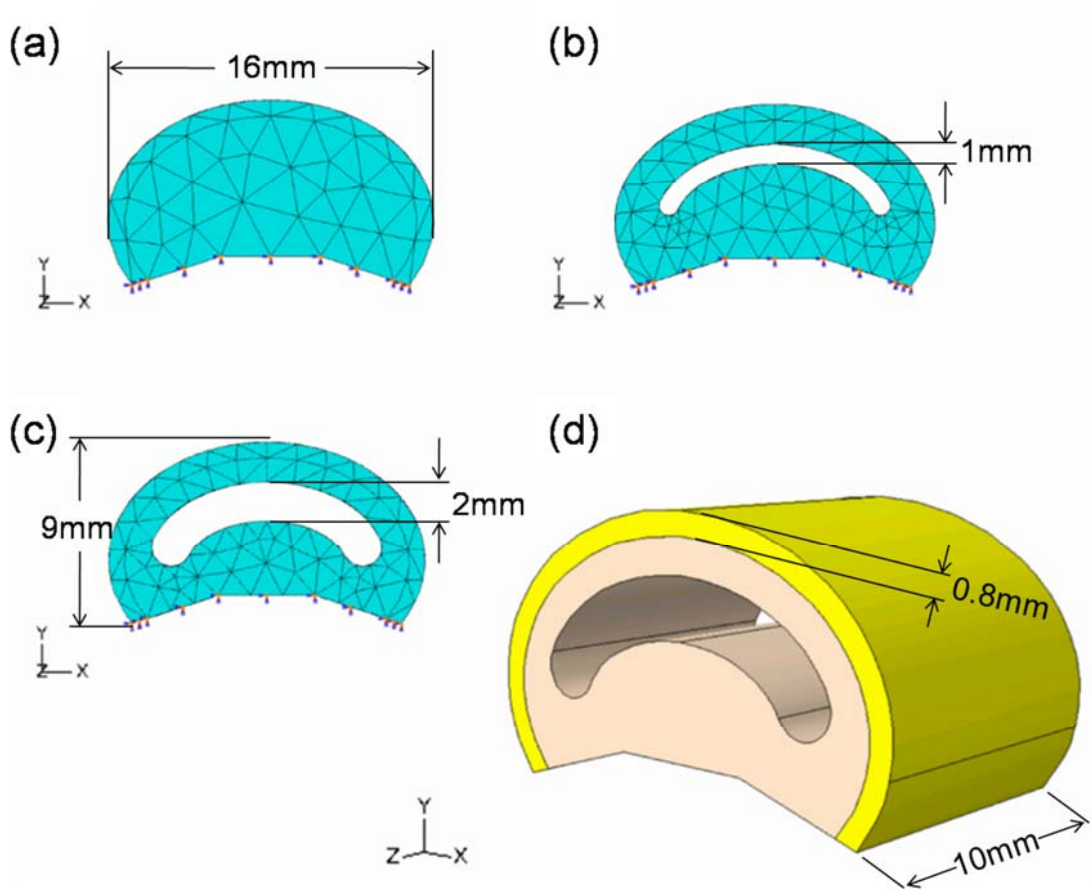

**Figure 3** – Geometries of the 3D finite element model. (a) Solid internal geometry. Internal geometry pockets of (b) 1 mm and (c) 2 mm heights. (d) The finite element model showing the two material layers.

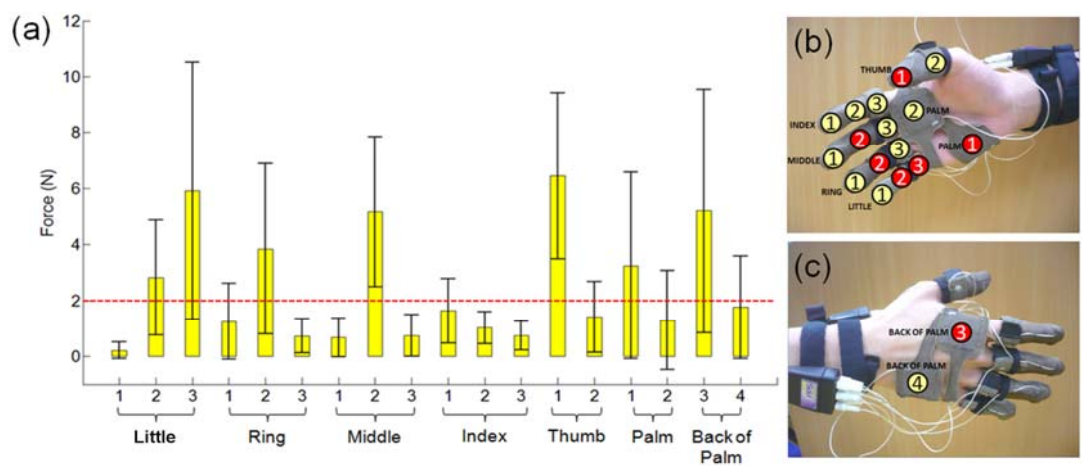

**Figure 4** – Parts of the hand with the corresponding contact forces during handshake. The data were taken from one male experimenter who shook hands with 30 male subjects. (a) Contact force distribution during handshake. The highlighted areas in red in (b) and (c) show the areas where contact forces are greater than 2 N.

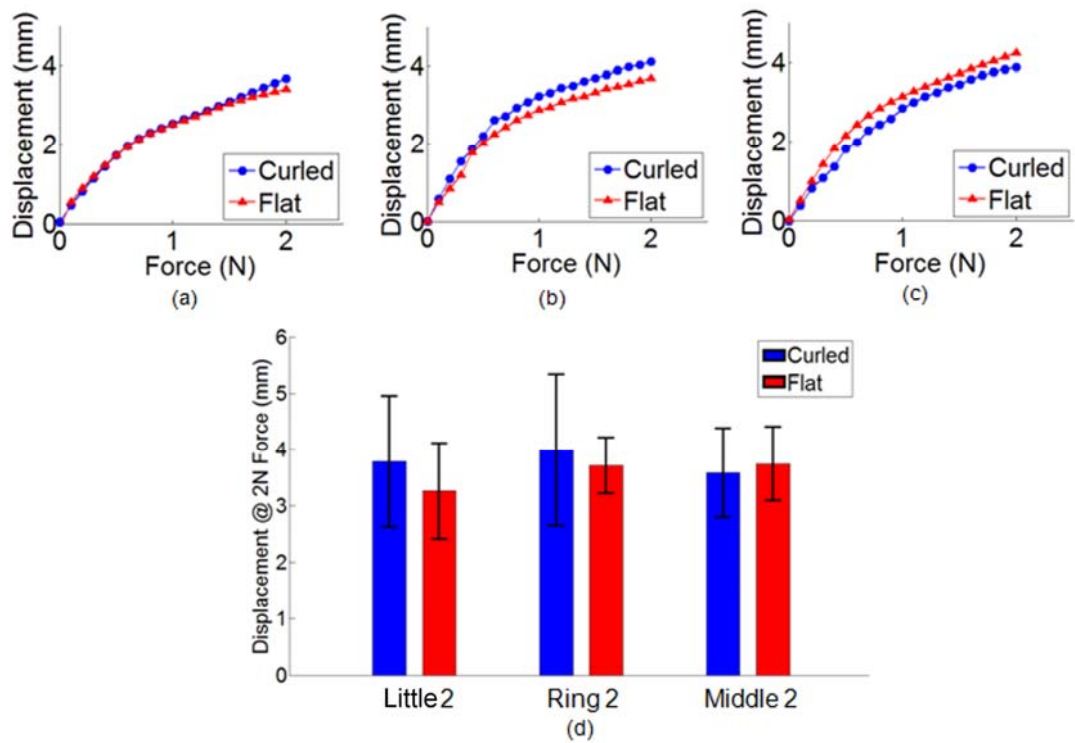

**Figure 5** – Indentation test results from one of the test subjects on a flat-hand position and a curled-fingers position at the (a) Little2, (b) Ring2, and (c) Middle2 finger phalanges. The bars in (d) show the comparison of the displacements at the Little2, Ring2, and Middle2 with 2 N force indentations for 10 subjects. The error bars represent the standard deviation.

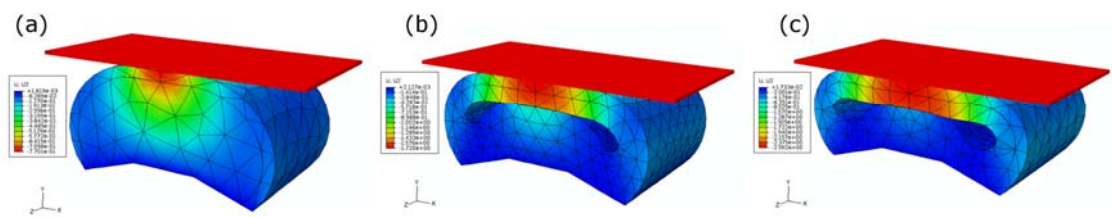

**Figure 6** – Finite element simulation results showing the displacement contours at 2 N force indentation. (a) Solid internal geometry. Internal geometry pockets with (b) 1 mm and (c) 2 mm heights.

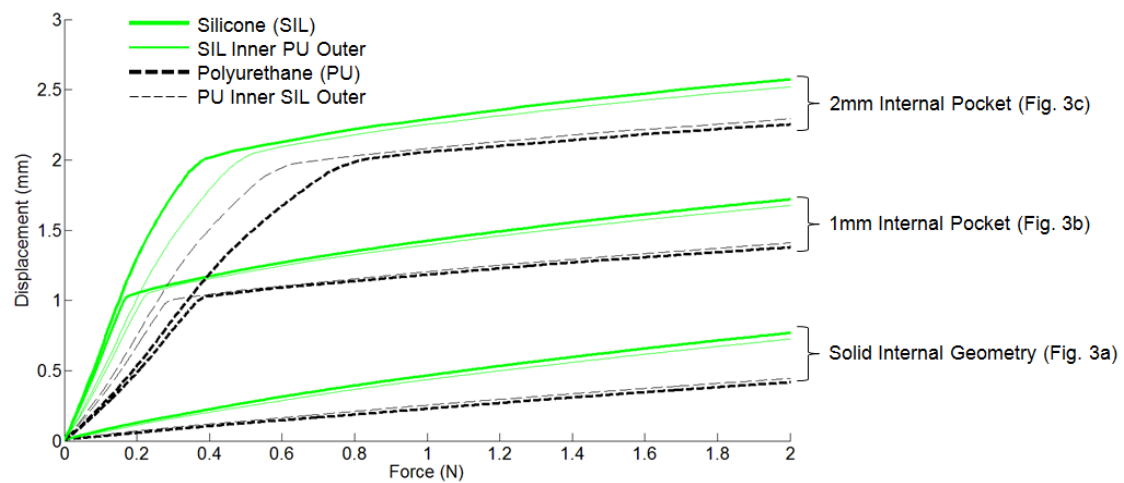

**Figure 7** – Finite element simulation results showing the effects of adding internal pockets to the synthetic skins (shown as thick lines) and the effects of varying the layers (shown as thin lines).

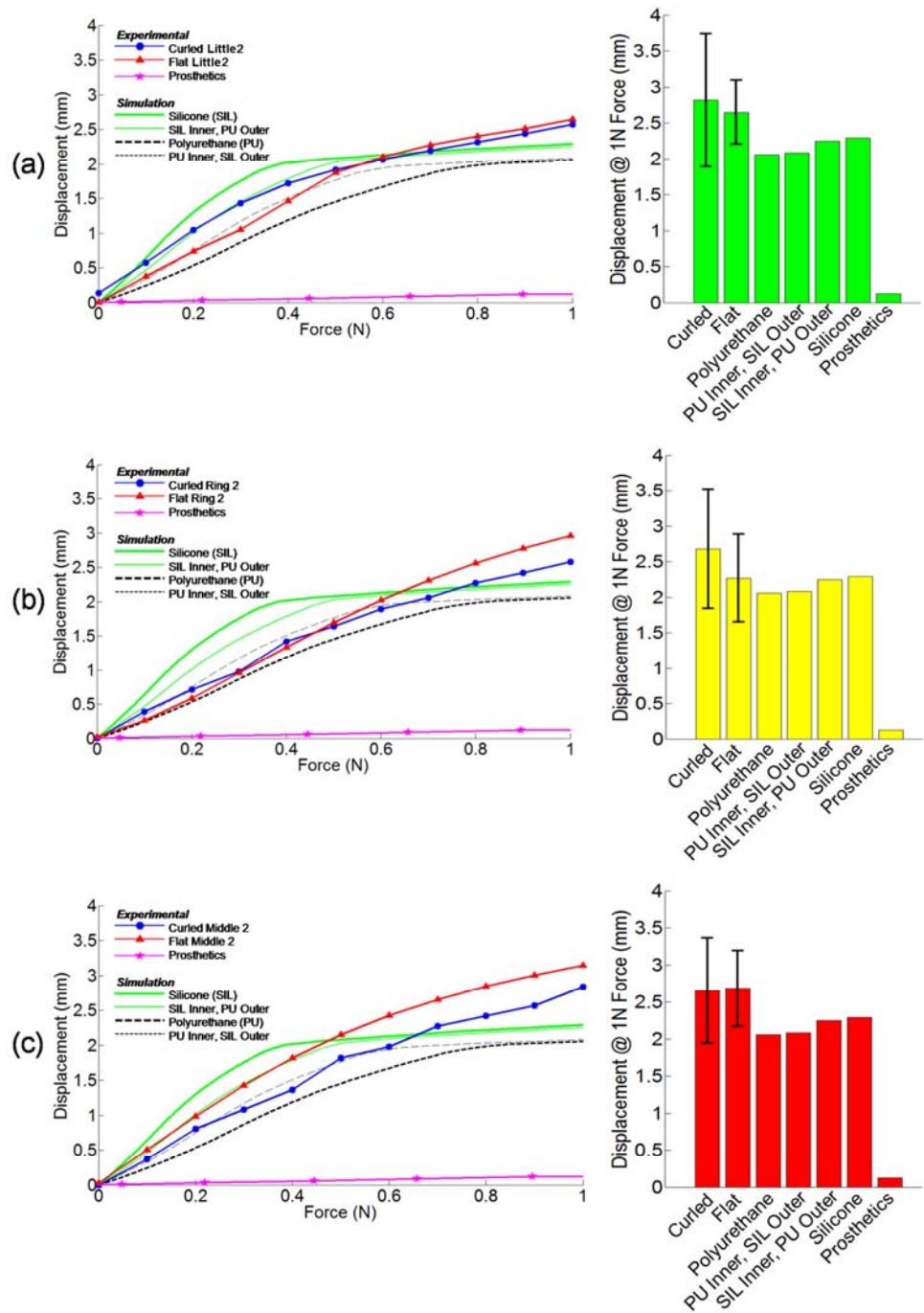

**Figure 8** – Comparisons of the finite element simulation results from the 2 mm inner pockets designs with the experimental results from the phalanx of a prosthetic hand and the human finger phalanges of 10 subjects on (a) Little2, (b) Ring2, and (c) Middle2 at 1 N force indentation.